\documentclass[journal,final,11pt,journal]{IEEEtran} 
\usepackage{cite}
\usepackage{amsmath,amssymb,amsfonts}
\usepackage{algorithmic}
\usepackage{graphicx}
\usepackage{textcomp}
\usepackage{algorithm}
\usepackage{mathrsfs}

\usepackage[T1]{fontenc}
\usepackage{balance}
\usepackage{soul}
\usepackage{algorithm}
\usepackage{algorithmic}
\usepackage{multicol}
\usepackage{multirow}
\usepackage{array}
\newcolumntype{P}[1]{>{\centering\arraybackslash}m{#1}}
\newcolumntype{a}{>{\columncolor{Gray}}c}
\usepackage{url}
\usepackage{makecell}
\usepackage[acronyms]{glossaries}
\newacronym{gat}{GAT}{Graph Attention Network}
\newacronym[plural=InstaGATs, firstplural=InstaGATs]{instagats}{\emph{InstaGATs}}{instantaneous GATs}

\newacronym{gnn}{GNN}{Graph Neural Network}

\newacronym{gru}{GRU}{Gated Recurrent Unit}

\newacronym{lstm}{LSTM}{Long Short-Term Memory}
\newacronym{lstmatt}{\emph{LSTM+Att}}{Long Short-Term Memory with Attention}

\newacronym{cnn}{CNNs}{Convolutional Neural Networks}
\newacronym{cnnatt}{\emph{CNNs+Att}}{Convolutional Neural Networks with Attention}

\newacronym{svm}{SVM}{Support Vector Machine}
\newacronym{da}{DA}{Discriminant Analysis}
\newacronym{lda}{LDA}{Linear Discriminant Analysis}
\newacronym{rf}{RF}{Random Forest}
\newacronym{nb}{NB}{Naive Bayes}
\newacronym{mlp}{MLP}{Multi Layer Perceptron}
\newacronym{rnn}{RNN}{Recurrent Neural Network}

\newacronym{sgd}{SGD}{Stochastic Gradient Descent}

\newacronym{cbam}{CBAM}{Convolutional Block Attention Module}

\newacronym{ica}{ICA}{Independent Component Analysis}

\newacronym{asr}{ASR}{Artifact Subspace Reconstruction}

\newacronym{ml}{ML}{Machine Learning}
\newacronym{dl}{DL}{Deep Learning}

\newacronym{eeg}{EEG}{Electroencephalography}
\newacronym{bci}{BCI}{Brain-Computer Interface}

\newacronym{ADHD}{ADHD}{Attention-Deficit Hyperactivity Disorder}

\usepackage[compact]{titlesec}

\usepackage[absolute]{textpos}
\usepackage{everyshi}

\begin{document}
\bstctlcite{IEEEexample:BSTcontrol}

\title{Comparison of Attention-based Deep Learning Models for EEG Classification}
\author{Giulia~Cisotto~\IEEEmembership{Member,~IEEE}, 
        Alessio~Zanga~$^{*}$, 
        Joanna~Chlebus~$^{*}$, 
        Italo~Zoppis,   
        Sara~Manzoni,   
        and~Urszula~Markowska-Kaczmar 
        \thanks{Manuscript received...}
        \thanks{Part of this work was also supported by MUR (Italian Minister for University and Research) under the initiative \emph{"Departments of Excellence"} (Law 232/2016).}
        \thanks{G.~Cisotto is with the Dept. of Information Engineering, University of Padova, Italy, with the National, Inter-University Consortium for Telecommunications (CNIT), Italy, and also with the National Center of Neurology and Psychiatry, Tokyo, Japan (email: giulia.cisotto.1@unipd.it).}
        \thanks{A.~Zanga, I.~Zoppis, and S.~Manzoni are with the Department of Computer Science, University of Milano-Bicocca, Milano, Italy (email: a.zanga3@campus.unimib.it, italo.zoppis@unimib.it, sara.manzoni@unimib.it)}
        \thanks{J.~Chlebus and U.~Markowska-Kaczmar are with the Department of Computational Intelligence, Wroclaw University of Science and Technology, Wyb. Wyspianskiego 27, Wroclaw, Poland (email: chlebusjoanna@gmail.com, urszula.markowska-kaczmar@pwr.edu.pl).}
        \thanks{$^{*}$ A.~Zanga and J.~Chlebus equally contributed to the manuscript.}
}

\maketitle

\begin{abstract}
%
Objective: To evaluate the impact on \gls{eeg} classification of different kinds of attention mechanisms in \gls{dl} models. 
Methods: We compared three attention-enhanced \gls{dl} models, the brand-new \glspl{instagats}, an LSTM with attention and a CNN with attention. We used these models to classify normal and abnormal (i.e., artifactual or pathological) \gls{eeg} patterns.
Results: We achieved the state of the art in all classification problems, regardless the large variability of the datasets and the simple architecture of the attention-enhanced models. We could also proved that, depending on how the attention mechanism is applied and where the attention layer is located in the model, we can alternatively leverage the information contained in the time, frequency or space domain of the dataset.
Conclusions: with this work, we shed light over the role of different attention mechanisms in the classification of normal and abnormal \gls{eeg} patterns. Moreover, we discussed how they can exploit the intrinsic relationships in the temporal, frequency and spatial domains of our brain activity.
Significance: Attention represents a promising strategy to evaluate the quality of the \gls{eeg} information, and its relevance, in different real-world scenarios. Moreover, it can make it easier to parallelize the computation and, thus, to speed up the analysis of big electrophysiological (e.g., \gls{eeg}) datasets.
\end{abstract}

\begin{textblock*}{17cm}(1.7cm, 0.5cm)
\noindent\scriptsize This work has been submitted to the IEEE for possible publication. Copyright may be transferred without notice, after which this version may no longer be accessible.\\
\textbf{Copyright Notice}: \textcopyright 2020 IEEE. Personal use of this material is permitted. Permission from IEEE must be obtained for all other uses, in any current or future media, including reprinting/republishing this material for advertising or promotional purposes, creating new collective works, for resale or redistribution to servers or lists, or reuse of any copyrighted component of this work in other works.
\end{textblock*}

\begin{IEEEkeywords}
attention mechanism, brain networks, CNN, deep learning, EEG, graph attention network, LSTM.
\end{IEEEkeywords}


\section{Introduction}\label{sec:intro}

\gls{eeg} is an electrophysiological technique that allows to acquire the electrical activity of the brain with a very high temporal resolution, in a non-invasive way and a relatively low cost. These are the reasons why \gls{eeg} has become very popular in many fields, ranging from medical diagnostics, pervasive healthcare, smart fitness, as well as gaming and \gls{bci}~\cite{Martiradonna2020, AESM2018, EEGgames, BCI-ICC2013}.
Across different applications, the detection of pathological \gls{eeg} patterns is still a challenging question.
Early approaches using \gls{ml} relied on intense pre-processing and the extraction of well-established engineered features~\cite{Amin, BCI-ICC2021}.
%
However, in the last decade, several \gls{dl} models have been also successfully proposed (see~\cite{Roy_2019} for a recent survey).
As a result, the challenge has moved from the development of relevant engineered features to the necessity of massive data collection, needed to effectively train optimal \gls{dl} models. In such increasing amount of data, identifying the most relevant information has become an important challenge.
\emph{Attention}~\cite{Cho2015}, one of the most recent and influential ideas in \gls{dl}, allows to easily embed external knowledge into a \gls{dl} model and to learn which portions of the data are relevant to the final output. This mechanism is expected to improve the \emph{explainability}~\cite{Adadi2018,XAI2019}, i.e., interpretability, of a \gls{dl} network, and to make it easier to introduce parallel computing.
In~\cite{song2018}, the authors showed how the exploitation of simple attention mechanisms can highly enhance the performance of a \gls{dl} model over a standard \gls{lstm}. In fact, the former could effectively model the correlations between different measurements in a lenghty clinical dataset through the so-called \emph{inter-attention}. On the other hand, \gls{lstm} is known to hardly manage long-term memory dependencies.
Therefore, in the last few years, a number of different attention strategies have been applied to \gls{eeg}-based recognition.
Phan et al.~\cite{Phan} proposed deep bidirectional \gls{rnn}s with attention mechanism for single-channel automatic sleep stage classification. Zhu et al.~\cite{Zhu_2020} describe how to perform automatic sleep staging recognition using a attention-enhanced \gls{cnn}.
Sha et al.~\cite{Sha_2017} developed a \gls{gru}-based \gls{rnn} with \emph{hierarchical attention} for mortality prediction.
A novel \emph{multi-view attention} network (MuVAN)~\cite{Yuan} has been shown to learn fine grained attentional representations from multivariate time series. MuVAN provides two-dimensional attention scores to estimate the quality of information of each view within different time-stamps.
Velivckovic et al.~\cite{Velivckovic2017} and Zoppis et al.~\cite{Zoppis2020} proposed two different stacked architectures made by an attention-enhanced \gls{gnn} to detect epileptic seizures in a multi-channel \gls{eeg} dataset.
In~\cite{Yuan_2018}, \emph{ChannelAtt} has been designed to jointly learn both multi-view data representations from a multi-channel \gls{eeg} dataset and their contribution scores to dynamically identify irrelavant channels in a seizure detection problem.
\emph{FusionAtt}~\cite{Ye_2019}, a deep fusional attention network, can learn channel-aware representations of multi-channel biosignals, and dynamically quantify the importance of each channel.
Cho et al.~\cite{cho2014} and Ma et al.~\cite{Ma_2019} introduced \emph{AttnSense}, a framework to combine attention mechanisms with \gls{cnn} and \gls{gru} in order to capture the dependencies of the sensed signals in both the spatial and the time domains.
In~\cite{song2018}, the authors designed \emph{SAnD} (i.e., Simply Attend and Diagnose), an architecture that employs a masked \emph{self-attention} mechanism, together with positional encoding and dense interpolation strategies for incorporating temporal order.
Self-attention is also employed in~\cite{wu2019} as alternative dynamic convolutions.
Finally, in~\cite{Wei}, \gls{cnn} encode frequency bands-related information and find the global temporal context, taking advantage of a \emph{multi-head} self-attention of a \emph{transformer model}.
While in the abovementioned literature the architectures with attention mechanisms have been compared to baseline, i.e. networks devoid of attention, in our work, we investigated the possibility to exploit the same attention strategy over different datasets. Moreover, we studied how attention is applied to different \gls{dl} models. To this aim, we re-implemented some of them.
We selected the most commonly used \gls{dl} models for \gls{eeg} recognition: \gls{cnn}, \gls{lstm}, and \gls{gnn}.
Each of them was enhanced by the introduction of attention and we evaluated its impact on the final classification outcome.
The relevance of the attention mechanisms used in the different \gls{dl} models has been discussed in the context of the classification of normal versus abnormal, either artifactual or pathological, \gls{eeg} segments across different datasets.

The paper consists of the following sections: Sec.~\ref{sec:preproc} is devoted to \gls{eeg} signal preprocessig and feature extraction; in Sec.~\ref{sec:models} we introduce the three attention-enhanced models we considered for our investigations. 
Sec.~\ref{sec:resmet} describes the research methodologies and Sec.~\ref{sec:results} provides the results of the evaluation of our models over $3$ public \gls{eeg} datasets. Finally, in Sec.~\ref{sec:discussion} we discuss them and the impact of attention in each single model. Sec.~\ref{sec:conclusions} concludes our work and identifies the most relevant future perspectives.

\section{Pre-processing and feature extraction}\label{sec:preproc}
\begin{figure*}[h]
	\centering
    \includegraphics[width=1\hsize, height=6cm]{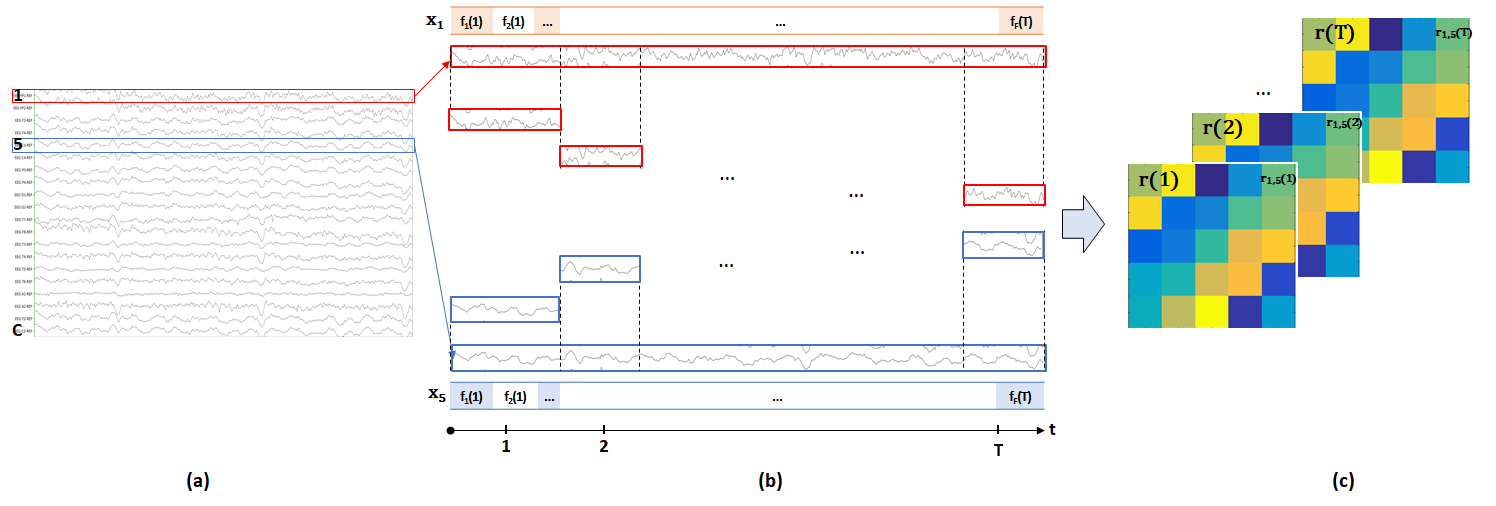}
    \caption{Data preparation (common for all models). (a) An example of a $C$-channels \gls{eeg} dataset. (b) Segmentation into $2$~s frames (in two representative channels, i.e., channel $1$ and $5$). (c) The set of $T$ correlation matrices, $r(t)$, obtained from each dataset with $T$ segments.}~\label{fig:dataprep}
\end{figure*}

In this work, we focused on \gls{eeg} data including normal and abnormal frames (i.e., segments), either pathological or artifactual.
%
Eleven well-established time-domain and frequency-domain features were computed from each EEG frame of each individual EEG channel. Particularly, in the time domain, we extracted the mean, the variance, the zero-crossings, the area under the curve, the skewness, the kurtosis, and the peak-to-peak distance (as listed in\cite{Zhang_2019}). Then, in the frequency domain, we computed the spectral power in well-known and clinically relevant frequency bands, i.e., the \emph{delta} ($\delta$) band corresponding to $(0.5,4)$~Hz, the \emph{theta} ($\theta$) band to $(4,8)$~Hz, the \emph{alpha} ($\alpha$) band to $(8,12)$~Hz, and the \emph{beta} ($\beta$) band to $(12,30)$~Hz.
In the following, we refer to the set of time- and frequency-domain, channel-wise, features as the vector: 
\begin{equation}
\mathbf{x_c}(t)=[f_1(t),f_2(t),...,f_{F}(t)],
\label{eq:feature_vector}
\end{equation}
with $F=11$ and $t = 1,2, ..., T$ where $T$ represents the number of available ordered time frames. 
Additionally, in each time frame $t$, we computed the Spearman's correlation coefficient between each pair of EEG channels obtaining a correlation matrix $\mathbf{r}$ for each available time frame $t$: 
\begin {equation}
\mathbf{r}(t) =   \left[
                    \begin{matrix}
                         r_{11}(t)  & r_{12}(t) & \cdots & r_{1C}(t) \\
                         r_{21}(t)  & \ddots    & \cdots & \vdots    \\
                         \vdots     & \vdots    & \ddots & \vdots    \\
                         r_{C1}(t)  & \cdots    & \cdots & r_{CC}(t) \\
                    \end{matrix}
                \right]
                \label{eq_correlation}
\end{equation}
with $t = 1,2, ..., T$, given $T$ the number of available frames and $r_{ij}(t)$ the correlation coefficient in the frame $t$ between channel $i$ and channel $j$.
%
At each time frame, the input to the deep learning models used in this work includes both the feature vector $\mathbf{x_c}$ for all channels $c=1,2, ..., C$, and the correlation matrix $\mathbf{r}$.
Unless explicitly specified, the dependence on the time frame, i.e., $t$, will be discarded to avoid confusion.
Fig.~\ref{fig:dataprep} reports an example of EEG recordings and all pre-processing steps to obtain the input to each deep learning model.

\section{Attention Models}~\label{sec:models}
The three attention-enhanced \gls{dl} models considered in this work share similar architectures that differ in the first layer.
\emph{\glspl{instagats}} (see Section~\ref{sec:instagats}) performs graph convolution on the input information; the \emph{\gls{lstm} with attention} model (see Section~\ref{sec:lstmatt}) exploits an \gls{lstm} at the first layer to process the time-related information in the input data; finally, the \emph{CNN with attention} model (see Section~\ref{sec:cnnatt}) performs a one-dimensional convolution operation on the input.
Next layers in each model have identical architecture: an LSTM layer, a dense layer, and a classification layer. The attention mechanism of each model fits the processing specificity in the model's first layer. The attention is placed between the model's first layer and the LSTM layer, except the LSTM network with attention, where it is located after the second LSTM layer.

We used categorical cross-entropy as a loss function for parameter optimization in all models, defined as follows:  
$$
\mathscr{L} = - \frac{1}{n}\sum_{i=1}^n \sum_{j=1}^m (y_{i,j}log(p_{i,j})) \label{eq:crossen}
$$
 where $y_{i,j}$ is the correct label class for the frame $i$, $p_{i,j}$ is the predicted output for it, $n$ is the number of samples in the dataset, $m$ is the number of classes. In our investigations, we considered only two classes and the output was one-hot encoded. 
The final layer of every model implemented the $softmax$ function.
To update parameters, mini-batch gradient descent method was applied. The latter updated the model parameters using a batch consisting of $B$ samples. The batch size $B$ was empirically set.

\subsection{InstaGATs}  \label{sec:instagats}
\begin{figure}[h]
\centering
    \includegraphics[width=0.5\textwidth, height=11cm]{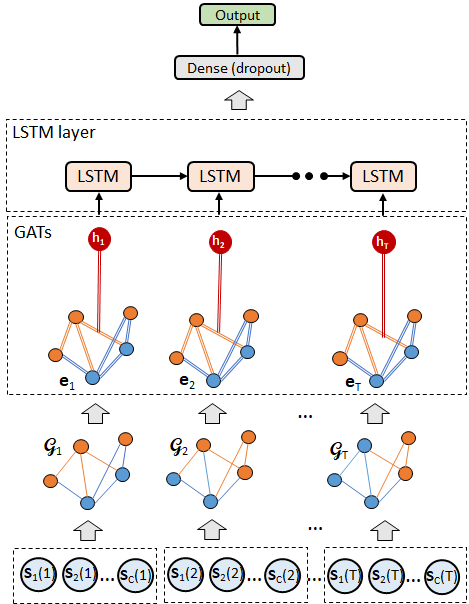}
    \caption{The architecture of the \gls{instagats} model (modified from~\cite{Zoppis2020}).} \label{fig:gat}
\end{figure}

Based on the seminal paper of~\cite{Velivckovic2017} implemented in~\cite{Zoppis2020}, a stacked architecture is proposed. The model is visualised in Fig.~\ref{fig:gat}.  It was obtained by placing an \gls{lstm} layer on the top of the \gls{gat} layer. The latter implemented attention and performed graph convolution. A dense layer with dropout and a $softmax$ layer completed the model architecture.
%
%
\gls{gat}~\cite{Velivckovic2017} operated on graphs; thus, it was necessary to represent the input data into a graph $\mathcal{G}$ at each $t$-th frame, $t\in \{1,2,..,T\}$. To this aim, at each $t$, we derived the adjacency matrix of graph $\mathcal{G}(t)$ from the correlation matrix $\mathbf{r}(t)$, as in \eqref{eq_correlation}. Thus, the \gls{gat}'s input matrix is formed by $C$ rows, each one defined by:
\begin{equation}
 \mathbf{s}_{i}(t) = [r_{1i}(t),r_{2i}(t),...,r_{Ci}(t)|| f_{1i}(t), f_{2i}(t), ... f_{Fi}(t)] \label{eq:concvec}
\end{equation}
where $r$ refers to the correlation coefficients in $\mathbf{r}(t)$, as in \eqref{eq_correlation}, $f$ to the time-domain and frequency-domain features, as in \eqref{eq:feature_vector}, and $||$ is the symbol of concatenation.
Each $\mathcal{G}(t)$ was formed by $C$ nodes (each corresponding to one \gls{eeg} channel) and $C^2$ edges.
Each node $i \in \{1,2,...,C\}$ was described, at time frame $t$, by the feature vector $\mathbf{x}_{i}(t)$, as in \eqref{eq:feature_vector}.
Given $\mathcal{G}(t)$, $t\in \{1,2,..,T\}$, \emph{\glspl{instagats}} performed graph convolution with attention on each node and its neighborhood (using their feature vectors) and provided the so-called \emph{embeddings}, $\bf{h_t}$, which were then delivered to the \gls{lstm} layer (Fig.~\ref{fig:gat}).

The attention mechanism was introduced by adapting the approach proposed in~\cite{Velivckovic2017}. With reference to \gls{gat}, attention defines the relevance of every particular feature in the node's feature vector during graph convolution. The latter is expressed by the attention coefficients. Formally, given $\mathcal{G}(t)$ with nodes' feature vectors $\bf{x}_i$, with $i \in \{1,2,...,C\}$, attention is obtained by evaluating the following function
$ a: \mathcal{R}^F \times \mathcal{R}^F \rightarrow \mathcal{R}$
which computes the coefficients $i_{v,u} = a (\bf W_G {x}_v, W_G \bf{x}_u)$ across every pair of nodes $(v,u)$, with  $(v,u) \in C \times C$, based on their feature vectors $\bf{x}_v$ and $\bf{x}_u$, respectively. $\bf W_G$ is a $F \times F$ matrix of weights between the nodes' features.
Thus, the attention coefficients $i_{v,u}$ quantify the significance of each feature for a node $u \in \mathcal{N}$, where $\mathcal{N}$ is in the first order neighborhood of node $v$. It is worth noting that all model's parameters, including the attention coefficients, were optimized, end-to-end, using the loss function described in \eqref{eq:crossen}.
%
%
By means of \emph{\glspl{instagats}}, we could capture both the relevant topology of the corresponding graph and the temporal dependency across different \gls{eeg} channels, while discarding irrelevant data from the \gls{lstm}'s memory.

\bigskip

\subsection{\gls{lstm} with Attention} \label{sec:lstmatt}
\begin{figure}[h]
 \centering
    \includegraphics[width=.4\textwidth, height=8cm]{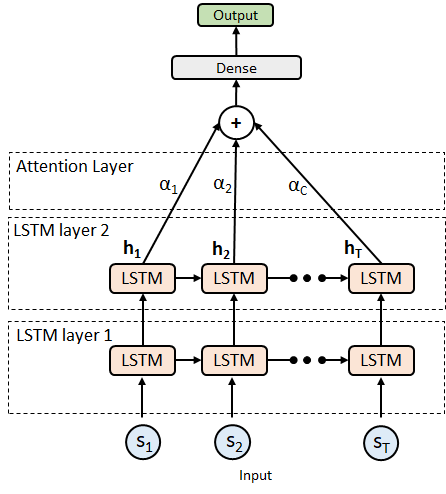} 
    \caption{The architecture of the \gls{lstmatt} (adapted from~\cite{Zhang_2019}).} \label{fig:LSTM} 
\end{figure}

As the next model, we referred to the network presented in~\cite{Zhang_2019}, and we implemented the \gls{lstmatt}.
Compared to its original version with 3-layer \gls{lstm} architecture, we focused on a 2-layer \gls{lstm}, to be consistent with the other models considered in this study.
On the top of the input \gls{lstm} layer with $T$ cells, we added another \gls{lstm} layer with $T$ \gls{lstm} cells, as in \emph{\glspl{instagats}}. Then, there is a dense layer delivering its information to the classification layer, consisting of $2$ neurons. The network was trained using the loss function described in \eqref{eq:crossen}. For each \gls{eeg} frame $t$, the input vector was created using the concatenation of the Spearman's correlation coefficients from $\mathbf{r}(t)$ and the vector of time-domain and frequency-domain features, $\mathbf{x_c}(t)$.
The concatenated vector for one frame is expressed by:
\begin{equation}
\mathbf{s}(t)=\left[\mathbf{s}_1||\mathbf{s}_2||..||\mathbf{s}_C\right] \label{eq:inputvec}
\end{equation}
where each $\mathbf{{s}_i}$ is described by \eqref{eq:concvec}. 
As depicted in Fig.~\ref{fig:LSTM}, the attention layer was stacked on the top of the second \gls{lstm} layer, assigning an appropriate weight $\alpha_{i}$ to each $i$-th cell output, $\bf{h}_i$, of the \gls{lstm} layer. Each vector $\bf{h}_i$ was multiplied by its weight $\alpha_{i}$ and, then, after concatenation of $T$ vectors, we obtained one vector, which was delivered to the dense layer without dropout. The \gls{eeg} classification was performed by the last layer implementing $softmax$.
Each \gls{lstm} layer's cell built its own representation of the input frame. We can say that, in this model, the attention mechanism focused on time-steps (frames) with the most discriminative information, assigning them higher coefficients $\alpha_{i}$.
More formally, if $\mathbf{h_i}$ was the output from the $i$-th cell of the second \gls{lstm} layer, the attention coefficients were calculated as $u_i=\tanh (\bf{W}_s \bf{h}_i)$, with $i \in \{1,2,...,T\}$ and $\bf{W}_s$ was a weight matrix. Next, they were normalized and the attention weights $\alpha_{i}=softmax(u_{i})$ were computed.
Unlike the original paper, we trained the model using the categorical cross-entropy loss function, as in \eqref{eq:crossen}.
%
%
\subsection{CNNs with Attention}  \label{sec:cnnatt}
\begin{figure}[h]
 \centering
    \includegraphics[width=.45\textwidth, height=9cm]{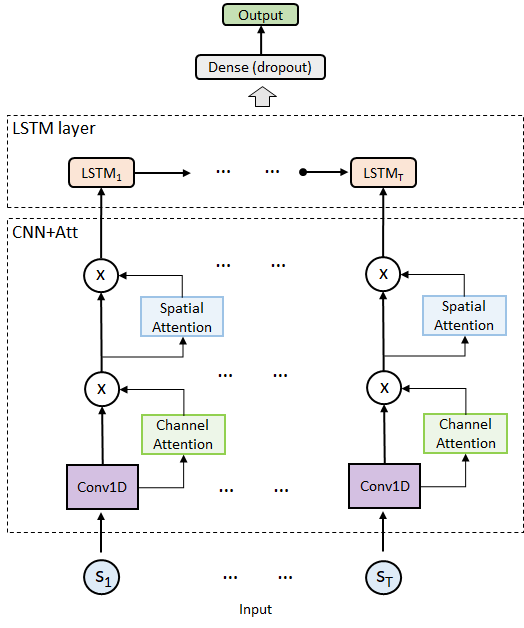} 
    \caption{The architecture of the \gls{cnnatt} model (adapted from~\cite{woo2018cbam}). }
    \label{fig:CNN}
\end{figure}

Lastly, we implemented a model based on \gls{cnnatt}. This model was adapted from~\cite{woo2018cbam}, where the authors introduced the \gls{cbam}, an attention mechanism specifically designed for convolutional networks.
The latter consisted of two kinds of attention mechanisms: a channel attention and a spatial attention sub-module. They worked complementary. The first one was responsible for defining \emph{what} was meaningful in the input, while the second focused on \emph{where} the relevant information was placed. Relevance was determined by the attention coefficients matrix $A_c$ for the channel attention (computed via shared \gls{mlp}) and $A_s$ for spatial attention (as the result of the convolutional operation). According to the authors' suggestion, we applied them sequentially, i.e., first the channel sub-module and, then, the spatial one.
As depicted in Fig.~\ref{fig:CNN}, the \gls{cnnatt} overall architecture is composed of a \gls{cnn} layer, combined with the \gls{cbam} module, an \gls{lstm} layer and a dense layer with dropout. 
The model performs 1D convolution operation on each input vector (defined as in \eqref{eq:inputvec}), representing a time frame (t), $t\in{1...,T}$.
A first refinement of the input feature map was obtained by multiplying it with the matrix $A_c$ (the \emph{channel attention map}); then, a second one via multiplication with $A_s$ (the \emph{spatial attention map}).
The outputs from the \gls{cnn} layer were concatenated and fed to the \gls{lstm} layer.
%
%
%
\subsection*{Baseline models without attention and implementation}
As a comparison for the models with attention previously described, we considered the following three models: a \gls{gnn}, an \gls{lstm} and a \gls{cnn}.
The latter share the same architectures with their respective attention-doped models, but the attention layer was removed.
All models were implemented in Python using Tensorflow 2 framework~\cite{Tensorflow}. For \gls{instagats} and the \gls{gnn}, the Spektral library of~\cite{grattarola2019} was used.
We optimized all hyper-parameters by selecting the values where the F1-score, as averaged across all datasets, was the highest.
Stochastic Adam optimizer~\cite{ADAM} was used for the models' parameters optimization.
Tab.~\ref{Tab:hyper_params} reports the optimized values for all models.

\section{Research methodologies and evaluation}~\label{sec:resmet}
In this section, we describe the public datasets we analyzed, the strategy for optimizing the hyper-parameters of the models, and we define the validation methodologies and metrics used to evaluate the different models.

\subsection{Datasets}\label{sec:datasets}
Three different EEG datasets which are provided by the Temple University Hospital of Philadelphia (Pennsylvania) were used in this work. All of them are publicly available, either on PhysioNet~\cite{physionet} or on the Temple University Hospital of Philadelphia (Pennsylvania) repository website~\cite{TUHEEGArtifact}. These datasets are slightly imbalanced with a ratio of $1:3$, in the worst case (see Tab.~\ref{Tab:sizes}). Then, we did not implement any balancing technique in our original datasets.

The first dataset is called \emph{TUH EEG Abnormal}~\cite{thesisAbnormal}.
Recording annotations are available, including patients' clinical history, medications, comments from expert visual inspection.
EEGs were recorded using a sampling frequency of 250 Hz and 16 bit resolution. Two montages, i.e., the averaged reference (AR) and the linked ears reference (LE), were used, depending on the file.
From this dataset we obtained $1472$ positive samples (with any kind of abnormality, for further details see \cite{thesisAbnormal}) and $1521$ negative samples (normal EEG samples).

The second dataset \emph{TUH EEG Artifact}~\cite{TUHEEGArtifact} contains both normal EEG signals and EEG signals affected by $5$ different types of artifacts: $161$ chewing events (\emph{chew}), $606$ eye movements (\emph{eyem}), $254$ muscular artifacts (\emph{musc}), $60$ shivering events (\emph{shiv}) and  $178$ instrumental artifacts (\emph{elpp}, e.g., electrode pop, electrostatic artifacts or lead artifacts).
Therefore, from this dataset we obtained $1259$ positive samples (with any kind of artifact, regardless their origin) and $1940$ negative samples (clean EEG samples).
EEGs were recorded using a sampling frequency of 250 Hz and 16 bit resolution.
To note, in the positive class we decided to include all kinds of artifact despite their different origin (e.g., physiological or instrumental) to train the models to recognize a good EEG pattern from an abnormal one. This is, in fact, in line with traditional EEG signal processing where identification and exclusion of artifactual EEG frames is the first - and, typically, time consuming - step of pre-processing.

The third dataset, named \emph{TUH EEG Seizure}~\cite{TUHEEGSeizure}, contains EEG recordings affected by different types of seizures.
Recordings include $19$ EEG channels and annotations by very well-trained medical doctors. \gls{eeg}s were recorded using a sampling frequency of 250 Hz and 16 bit resolution with a bipolar channel configuration.
From this dataset, we extracted $4240$ samples with focal non-specific seizures and $1717$ samples with generalized non-specific seizures.
Focal non-specific seizure samples were included in the \emph{negative} class, while the generalized non-specific seizure samples in the \emph{positive} class. This choice reflects the relative severity of one type of seizure compared to the other: indeed, focal seizures affect a limited portion of brain activity compared to the generalized ones. The latter, thus, possibly lead to more severe brain damages.

The datasets significantly differ by size, study aim as well as number of events and even \gls{eeg} acquisition setup. However, this variety allowed us to widely assess the performance of different \gls{dl} models, evaluating their ability to classify \gls{eeg} in several specific, but different, classification problems.

\subsection{Data preparation}
Despite the heterogeneity of the datasets, a few common steps of pre-processing were applied: the \gls{eeg} recordings from every channel (i.e., sensor location) were filtered in the frequency band $(0.1, 47)$~Hz, normalized using a min-max centered normalization technique~\cite{Basheer2000} and segmented into $2$~s frames using a sliding window approach, in line with the most common literature on \gls{eeg} (e.g., see~\cite{Healthcom2018,cisotto2020}).
The sampling frequency and the number of channels could differ from dataset to dataset, and even from file to file. Therefore, as a first step, we downsampled all of them to the lowest available sampling frequency, i.e., $250$~Hz. Then, the largest common set of $C$ channels for all files was retained. The pre-processing phase and the data preparation phase were carried out using \emph{PyEEGLab}~\cite{Zanga2019PyEEGLab}. 
%
After band-pass filtering and segmentation (see Section~\ref{sec:preproc}), we obtained a roughly balanced number of frames per class in each dataset, as reported in Table~\ref{Tab:sizes}. 
\begin{table}[htb]
\centering
\caption{Number of positive and negative instances in each dataset.} \label{Tab:sizes}
\setlength{\tabcolsep}{3pt}
\begin{tabular}{|p{100pt}|p{60pt}|p{60pt}|}
\hline
\textbf{Dataset name} & \textbf{Positive class}  & \textbf{Negative class}  \\
\hline
TUH EEG Abnormal & $1472$       & $1521$ \\ 
\hline
TUH EEG Artifact & $1259$  & $1940$ \\
\hline
TUH EEG Seizure  & $1717$       & $4240$  \\
\hline
\end{tabular}
\end{table}

\begin{table*}[htb]
\centering
\caption{Optimzed hyper-parameter values for every model.}
\label{Tab:hyper_params}
\setlength{\tabcolsep}{3pt}
\begin{tabular}{|p{100pt}|p{100pt}|p{60pt}|p{40pt}|}
\hline
\bf{Parameter}        & \bf{Range} & \textbf{\gls{instagats}} & \textbf{GNN}  \\
\hline
GATs output channels  & $(8,64)$          & 64                & 32            \\
LSTM hidden units     & $(8,256)$         & 64                & 64            \\
Dropout               & $[0, 0.2]$        & 0.2               & 0.15          \\
Learning rate         & $\{10^{-4}, 5~10^{-4}, 10^{-3}\}$  & 0.0005            & 0.0001        \\
\hline
\bf{Parameter}        & \bf{Range}       & \textbf{\gls{lstmatt}} & \textbf{LSTM} \\
\hline
LSTM hidden units     & $(8,256)$        & 128               & 128           \\
LSTM L2 reg           & $[0.001, 0.04]$ & 0.001             & 0.001         \\
Input dropout         & $[0, 0.2]$       & 0.1               & 0.1           \\
LSTM layer 1 dropout  & $[0, 0.4]$       & 0.2               & 0.2           \\
LSTM layer 2 dropout  & $[0, 0.4]$       & 0.2               & 0.2           \\
Learning rate         & $\{10^{-4}, 5~10^{-4}, 10^{-3}\}$ & 0.0001            & 0.0001        \\
\hline
\bf{Parameter}                   & \bf{Range}       & \textbf{\gls{cnnatt}}  & \textbf{CNN}  \\
\hline
Conv kernel                      & $\{3,5,7,11\}$   & 3                 & 3             \\
Conv filters                     & $\{4,8,16,32\}$  & 32                & 8             \\
LSTM hidden units                & $(8,256)$        & 256               & 8             \\
Dropout                          & $[0, 0.5]$       & 0.15              & 0.15          \\
Learning rate                    & $\{10^{-4}, 5~10^{-4}, 10^{-3}\}$ & 0.001             & 0.001         \\
CBAM reduction ratio             & $\{4,8,16\}$     & 16                & -             \\
CBAM spatial kernel              & $\{5,7,11\}$     & 7                 & -             \\
\hline
\end{tabular}
\end{table*}

%
%


\subsection{Evaluation and performance metrics}\label{sec:performance_metrics}
To increase the reliability of results and to obtain a more robust error estimation, a stratified $10$-fold cross-validation was applied. In each fold, every model was trained for $50$ epochs with a batch size of $32$. In order to evaluate the models performance, we used well-established classification metrics: accuracy, recall, precision and F1-score. Recall and precision are the most relevant metrics to investigate the effectiveness of a classifier when  searching for rare, but pathological, samples. 
Given $N$ as the total number of samples to classify, $TP$ (True Positive) the number of samples correctly classified as positive, $TN$ (True Negative) the number of samples correctly classified as negative, $FP$ (False positive) the number of samples incorrectly classified as positive, and $FN$ (False Negative) the number of samples incorrectly classified as negative, we could calculate the accuracy, recall, precision, and F1-score.

\section{Results}~\label{sec:results}
\begin{figure*}[h]
    \centering
    \includegraphics[width=0.3\textwidth,height=3cm]{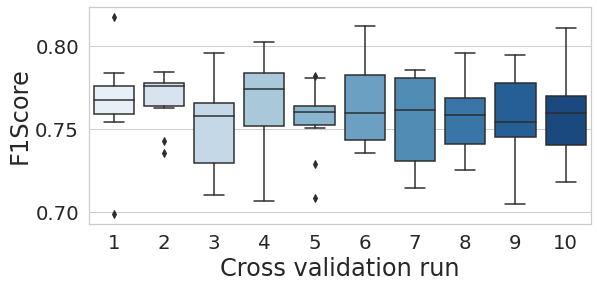}
    \includegraphics[width=0.3\textwidth,height=3cm]{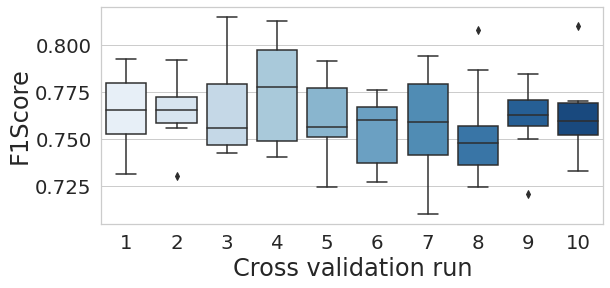}
    \includegraphics[width=0.3\textwidth,height=3cm]{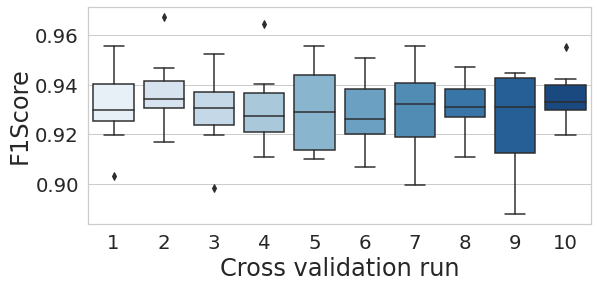}
    \caption{F1-scores for \gls{instagats} in cross-validation. (a) \emph{TUH EEG Abnormal}. (b) \emph{TUH EEG Artifact}. (c) \emph{TUH EEG Seizure}.}  
    \label{fig:instaGAT boxplots}
\end{figure*}

%
In this research, we evaluated the performance of $3$ models with attention, as described in Section~\ref{sec:models}, in $3$ very different classification problems over \gls{eeg}, with the aim to evaluate to what extent similar architectures with attention can classify different kinds of \gls{eeg} patterns.
Tab.~\ref{Tab:results} reports all results in terms of accuracy, recall, precision and F1-score, for every model and every dataset. 
%
%
First, we used the models to identify those \gls{eeg} segments which included any abnormal spike, K-complex or any other pathological waveform in the \emph{TUH EEG Abnormal}.
Here, the best accuracy performance was scored by the baseline \gls{lstm} model with $79.18\%$, while the best F1-score was achieved by \emph{\gls{lstm}$+$Att} model with $78.88\%$. However, most models provided an accuracy and F1-score values above $75\%$, except for the \gls{cnn} model which reached lower values (around $68\%$).
%
In \emph{TUH EEG Artifact} dataset, we aimed to detect \gls{eeg} segments where any kind of physiological or instrumental artifact was present. 
Again, \gls{lstm} and \emph{\gls{lstm}$+$Att} obtained the best results.
However, from Tab.~\ref{Tab:results}, we noticed that \gls{instagats} and \emph{\gls{cnn}$+$Att} achieved significantly better results with respect to their baseline models without attention.
%
Finally, in \emph{TUH EEG Seizure}, we reached the best accuracy ($96.98\%$) and F1-score ($94.71\%$) with the \gls{cnnatt}.
However, here, all models obtained comparable high accuracies (above $95\%$) and F1-scores (above $92\%$).
%
To note, when comparing with the state of the art, we typically focus on accuracy, as the most common classification metric used in the literature. However, since the datasets are not fully balanced and we need to keep into consideration pathological events (i.e., relevant positive samples), the F1-score has a key role to take into account two different kinds of error that can occur: type I, or false positive samples, and type II, or false negatives.
Both of them are significant when dealing with pathological \gls{eeg}. However, the desirable output from any classifier is to have high rates of correct detection of positive samples (i.e., recall), in order to identify the most abnormal \gls{eeg} segments, and to ensure low rates of false detection of positive samples (i.e., precision), in order not to mistakenly intervene on healthy regions of the brain.
\begin{table}[htb]
\centering
\caption{Classification results (mean and standard deviation reported for each metric)}
\label{Tab:results}
\setlength{\tabcolsep}{3pt}
\begin{tabular}{lrrrrrrrr} 
\hline
\multicolumn{1}{c}{} &
  \multicolumn{2}{c}{\bf{Accuracy}} &
  \multicolumn{2}{c}{\bf{Recall}} &
  \multicolumn{2}{c}{\bf{Precision}} &
  \multicolumn{2}{c}{\bf{F1-score}} \\ \cline{2-9} 
\multicolumn{1}{c}{Model} &
  \multicolumn{1}{c}{Mean} &
  \multicolumn{1}{c}{Std} &
  \multicolumn{1}{c}{Mean} &
  \multicolumn{1}{c}{Std} &
  \multicolumn{1}{c}{Mean} &
  \multicolumn{1}{c}{Std} &
  \multicolumn{1}{c}{Mean} &
  \multicolumn{1}{c}{Std} \\ \hline
\multicolumn{9}{c}{\bf{TUH EEG Abnormal}}                                 \\ \hline
\gls{instagats} & 77.25 & 1.86 & 76.70 & 3.48 & 77.18 & 3.50 & 76.83 & 1.66 \\
GNN             & 74.01 & 1.26 & 73.03 & 3.43 & 73.88 & 1.55 & 73.40 & 1.70 \\
\gls{lstmatt}   & 79.05 & 2.18 & \bf{79.48} & \bf{3.55} & 78.53 & 4.00 & \bf{78.88} & \bf{1.81} \\
LSTM            & \bf{79.18} & \bf{1.52} & 77.85 & 4.06 & \bf{79.56} & \bf{2.67} & 78.59 & 1.78 \\
\gls{cnnatt}    & 76.51 & 1.63 & 74.32 & 2.87 & 77.11 & 1.65 & 75.67 & 1.82 \\
CNN      & 68.29 & 1.29 & 62.64 & 3.70 & 69.84 & 1.60 & 65.98 & 1.99 \\ \hline
\multicolumn{9}{c}{\bf{TUH EEG Artifact}}                                 \\ \hline
\gls{instagats} & 81.40 & 1.42 & 76.81 & 3.46 & 76.34 & 3.80 & 76.47 & 1.61 \\
GNN             & 76.84 & 1.42 & 68.70 & 3.98 & 71.49 & 2.60 & 69.98 & 2.01 \\
\gls{lstmatt}   & 82.84 & 2.31 & \bf{79.51} & \bf{3.99} & 77.69 & 3.97 & \bf{78.49} & \bf{2.64} \\
LSTM            & \bf{83.15} & \bf{2.00} & 77.84 & 2.90 & \bf{79.23} & \bf{3.90} & 78.45 & 2.26 \\
\gls{cnnatt}    & 79.18 & 1.31 & 72.12 & 4.00 & 74.28 & 1.73 & 73.12 & 2.14 \\
CNN             & 72.18 & 2.71 & 64.90 & 4.20 & 64.61 & 3.39 & 64.72 & 3.53 \\ \hline
\multicolumn{9}{c}{\bf{TUH EEG Seizure}}                                  \\ \hline
\gls{instagats} & 96.29 & 0.54 & 92.49 & 1.78 & 94.55 & 1.40 & 93.49 & 0.97 \\
GNN             & 95.65 & 0.48 & 91.03 & 1.92 & 93.76 & 1.79 & 92.35 & 0.84 \\
\gls{lstmatt}   & 96.79 & 0.72 & 94.23 & 1.63 & 94.64 & 1.46 & 94.43 & 1.26 \\
LSTM            & 96.79 & 0.53 & \bf{94.41} & \bf{1.17} & 94.49 & 1.44 & 94.44 & 0.90 \\
\gls{cnnatt}    & \bf{96.98} & \bf{0.61} & 94.00 & 1.92 & \bf{95.48} & \bf{1.63} & \bf{94.71} & \bf{1.08} \\
CNN             & 95.70 & 0.82 & 91.73 & 1.31 & 93.32 & 2.71 & 92.49 & 1.35 \\ \hline
\end{tabular}
\end{table}
In Fig.~\ref{fig:instaGAT boxplots}, we also provide a representative boxplot to show the high variability (i.e., standard deviation) of F1-scores across datasets. It reports the scores for \gls{instagats} (similar tendency was observed for the other models, not included for space limitations).

\section{Discussion}~\label{sec:discussion}
The application of \gls{dl} models on real-world data, such as \gls{eeg} recordings, still represents a challenging task. Multiple concurrent factors, like technical equipment, recording noise, physical and emotional states are combined together in these datasets to form complex scenarios, in which inter- and intra-subject differences could strongly affect the classification performance.
In line with this observation, here we implemented and compared $3$ attention-doped \gls{dl} models (and their corresponding models without attention) over $3$ very different \gls{eeg} datasets, in order to investigate how attention can help \gls{dl} models to recognize abnormal \gls{eeg} patterns, generalizing the detection to both pathological and artifactual abnormalities. 
%
%
We showed that, in all classification problems, we reached the state of the art with most of the models (both with and without attention). In the \emph{TUH EEG Abnormal} dataset, the \gls{lstm} model achieved an accuracy of $79\%$ that is comparable with the models proposed in~\cite{lopez2015automated} (the authors collected, themselves, the dataset) which provided a variable error detection level between $21\%$ and $41\%$.
Also, the \emph{TUH EEG Seizure} dataset has been analyzed in other papers,~\cite{Li2020} among others, with accuracy values between $79\%$ and $92\%$ and F1-score between $75\%$ and $94\%$ (actually, we slightly outperformed the best literature results with all our models).
Finally, the \emph{TUH EEG Artifact} dataset has been considered by~\cite{Kim2019, Roy2019, Saeed2020}. They achieved a peak accuracy value of $74.99\%$ with a deep CNN model, $71.43\%$ with a standard \gls{lda} model, and $71.1\%$ with a CNN-based attention-enhanced architecture, respectively.
%
Moreover, we noticed that it seems to be easier to classify different kinds of seizures, i.e., focal versus generalized, compared to abnormal or other pathological events. This might be due to the well defined difference between the two classes. In fact, focal (non-specific) seizures are typically defined as short-lasting, space-limited seizure events, while generalized (non-specific) seizures spread across several brain regions (i.e., EEG sensors) at the same time and they are longer lasting.
On the other hand, the positive classes in the other two datasets, i.e., \emph{TUH EEG Abnormal} and \emph{TUH EEG Artifact}, contain a large variety of abnormal or artifactual events that are, possibly, harder to be accurately detected.
This might have facilitated all models to classify, with better performance metrics, seizures in the \emph{TUH EEG Seizure} dataset, compared to other events in the remaining datasets. Also, this might explain why results for \emph{TUH EEG Seizure} showed acceptable (small) standard deviations of the F1-scores in cross-validation, while in the other datasets variability is larger. 
%
%
Then, in order to evaluate the impact of the attention mechanisms in each proposed model, we point out that they were purposely simplified.
Their architectures were composed of one attention layer, which encoded a model-specific attention mechanism, an \gls{lstm} layer, and a dense layer that provided the output. This simple, but effective, common design enabled us to compare the different attention-enhanced models.
In fact, each attention mechanism was designed to exploit the input features with a specific representation, i.e., in the time, frequency and spatial domains. For example, the \gls{gat} model applied attention to a spatial representation derived from the first-order neighborhood's hierarchical structure. The mechanism leveraged the local spatial embeddings obtained from the input correlation matrix and the features of the nodes from neighborhood.
The \emph{\gls{lstm}+Att} model filtered out irrelevant information by applying attention in the temporal dimension.
Finally, the \gls{cbam} module of \emph{\gls{cnn}+Att} exploited attention for each \gls{eeg} channel, separately.
Following this approach, those models that primarily use spatial features showed a performance increase when attention was applied (e.g., \gls{instagats} over \gls{gnn}, and \gls{cnnatt} over CNN). Whereas, those architectures that rely on temporal features demonstrated no advantages when attention is added. 
Interestingly, the proposed attention-enhanced models could take advantage of different \gls{eeg} descriptions which can alternatively, or jointly, consider time, frequency and space (i.e., sensors location). In fact, such domains usually provide key pieces of information to interpret or analyze brain activity corresponding to different individuals' behaviours or tasks performance.
These considerations could prospectively provide important indications to setup appropriate experimental protocols and data processing pipelines, depending on what individual behaviour/task has to be investigated.
For instance, if specific brain regions interactions are expected, as in many cognitive tasks~\cite{REPAC}, an architecture such as \gls{instagats} could identify clusters of \gls{eeg} sensors that mostly account for the synchronization between those regions during the accomplishment of the cognitive task~\cite{REPAC}. In this case, it could be possible to leverage the \gls{instagats}'s spatial-dependant embeddings to identify the most involved brain areas.
%
Also, in case of reaction tasks~\cite{Chowdhury2020} (when the individual is expected to promptly react to an external stimulus), time-dependant features could be filtered out more accurately by an architecture such as \gls{lstmatt}.
%
%
%
We can also highlight that, despite their simple architectures, the attention mechanism allowed the models to achieve high accuracy levels across different real-world scenarios with minimal pre-processing.
The latter is very often guided by experts or based on field knowledge. However, this might lead to non-reproducible results over the same dataset, depending on the analyst who performs data analysis.
Thus, reducing pre-processing could represent an important advantage compared to standard \gls{ml} or to other \gls{dl} methods. 
%
This work still lacks a proper investigation of larger datasets affected by artifacts, where pre-processing might be highly needed. However, it might open the way to further investigations with the aim to reduce empirically-based pre-processing in big \gls{eeg} datasets.

\section{Conclusions}~\label{sec:conclusions}
%
In this paper, we aimed to evaluate the impact of different kinds of attention mechanisms, when added on well-established \gls{dl} models. We compared $3$ architectures: the brand-new \emph{\glspl{instagats}}, the \emph{\gls{lstm}+Att}~\cite{Zhang_2019} and a \emph{\gls{cnn}+Att}~\cite{woo2018cbam}.
We used these models to classify different kinds of \gls{eeg} normal and abnormal patterns, with further distinction between artifactual and pathological abnormalities. The \gls{eeg} datasets were available online and included a large variety of spikes, seizures, instrumental artifacts and many other relevant events.
Our results showed that we achieved the state of the art in all classification problems, regardless the large variability of the datasets and the simple architecture of the proposed attention-enhanced models.
We also observed that each kind of attention mechanism can enhance the identification of a specific \gls{eeg} domain characterization: e.g., in \emph{TUH EEG Artifact}, attention might have particularly helped \emph{\glspl{instagats}} and \emph{\gls{cnn}+Att}, compared to their corresponding models without attention, to identify artifacts, which typically spread across different brain areas. 
%
%
More generally, we showed that, depending on how the attention mechanism is applied and where the attention layer is located in the \gls{dl} model, we can alternatively leverage the information contained in the data in one of the above mentioned domains.
%
Interestingly, this can pave the way to further investigations on big \gls{eeg} datasets and on a more robust selection of the information provided by large amounts of data, in order to relate brain activity with the individuals' behaviour or task performance.
%
Therefore, attention represents a promising strategy to evaluate the quality of the \gls{eeg} information, and its relevance, in different real-world scenarios. Moreover, attention can make it easier to parallelize the computation and, thus, to speed up the analysis of big electrophysiological (e.g., \gls{eeg}) datasets.



\bibliographystyle{IEEEtran}
\bibliography{attentioneeg_references}

\end{document}